# Students' Information Privacy Concerns in Learning Analytics: Towards Model Development

Chantal Mutimukwe[1], Jean Damascene Twizeyimana[2] and Olga Viberg[1]

[1] Royal Institute of Technology (KTH), Sweden
chantalm@kth.se
oviberg@kth.se

[2] University of Rwanda, Kigali, Rwanda
tdamas3d@gmail.com

**Abstract.** The widespread interest in learning analytics (LA) is associated with increased availability of and access to student data where students' actions are monitored, recorded, stored and analyzed. The availability and analysis of such data is argued to be crucial for improved learning and teaching. Yet, these data can be exposed to misuse, for example, to be used for commercial purposes, consequently, resulting in *information privacy concerns* (IPC) of students who are the key stakeholders and data subjects in the LA context. The main objective of this study is to propose a theoretical model to understand the IPC of students in relation to LA. We explore the concept of IPC as a central construct between its two antecedents: perceived privacy vulnerability and perceived privacy control, and its consequences, trusting beliefs and self-disclosure behavior. Although these relationships have been investigated in other contexts, this study aims to offer mainly theoretical insights on how these relationships may be shaped in the context of LA in higher education. Understanding students' IPC, the related root causes and consequences in LA is the key step to a more comprehensive understanding of privacy issues and the development of effective privacy practices that would protect students' privacy in the evolving setting of data-driven higher education.
.

**Keywords:** Information Privacy Concerns, Learning Analytics, Students, Higher Education, Model Development.

# 1 Introduction

Learning Analytics (LA) refers to "the measurement, collection, analysis and reporting of data about learners and their contexts, for purposes of understanding and optimizing learning and the environments in which it occurs" [1, p.34]. Although the practices of data collection, analysis, and reporting are beneficial for improved teaching and learning [2], they are also associated with the risks of privacy violation of students' personal information [3]–[5]. One of the most critical issues is that students' personal information can be accessed and used for commercial purposes without students' awareness and/or consent [6]. Based on that, scholars highlight the need for further research that takes into account information privacy concerns of students in relation to LA [3], [7].

*Information privacy concerns* (IPC) have been identified as a major and central construct in a number of studies that have attempted to conceptualize *'information privacy'* in different contexts, including social media, commerce, governance, and health care contexts (e.g., [9]–[11]). IPC refer to individual concerns about the possible loss of privacy as a result of information disclosure to a specific external agent/institution [12]. Such concerns may vary from the intrusion of an individual's privacy to potential breaches that can lead to identity theft [13]. IPC reflect an individual's perception of his/her concerns/worry for how his/her personal information is handled by a specific institution, and this is different from his/her expectations, general perceptions, or awareness of how the institution should handle his or her personal information [10].

Researchers in information systems have earlier identified a number of factors, including awareness of information collection, unauthorized access, perceived vulnerability to information misuse, experience with Internet use, cultural background of users to be antecedents to individuals' IPC [14]–[17]. The consequences of such concerns could range from individuals declining or refusing to disclose personal information, and/or mistrust in online services [18], [19]. In the context of LA, all these consequences can slow down and/or hinder the adoption of LA, at scale. Tsai et al. [20] stress that "the associated [with LA promises to support adaptive learning in higher education] issues around privacy protection, especially their implications for students as data subjects, has been a hurdle to wide-scale adoption" [p. 230].

As mentioned above, there is a significant number of studies that have explored IPC in different contexts (e.g., e-commerce, e-government, e-health, social media. However, our analysis of extant literature shows this matter has so far received little attention in the context of LA. Studies that explored students' privacy – sometimes intermingled with other ethical issues from the student perspectives in LA – focused on expectations (e.g., [21], [22]), perceptions (e.g., [3]), awareness (e.g., [23]), and preferences (e.g., [24]). ToTo the best of our knowledge, few studies developed models and/or carried empirical studies to explain the construct of IPC, by together exploring its antecedents, and consequences.

Researchers suggest that to better address the problems of information privacy within a certain context, it is crucial to first comprehend the nature of IPC, its root causes, and consequences from data subject perspective in that particular context (e.g., [17], [18]). Moreover, others posit that IPC is context-specific, and it is important to understand how they may vary from one context to another or may be shaped by the type of information, the level of sensitivity, and the types of institutions and data collection and analysis practices ([9], [19]).

To address the gap in the existing research, this study aims at proposing a model that would explain the construct of IPC from the student – the primary data subjects – perspective in the LA context. We explore it as a central construct, between its two antecedents; *perceived privacy risks* and *perceived privacy control*, and its consequences, *trusting beliefs*, and *self-disclosure behavior*. For this study, we consider students from high education institutions since the prevailing part of the LA research has hitherto been performed in the setting of higher education [2]. Yet, Slade et al. [8] pointed out that higher education institutions are increasingly making a complex, but fluid context where governments, business entities, and data brokers can collect, analyze and exchange information.

## 2  Research background

### 2.1  Defining privacy

Privacy is a complex concept that is associated with several interrelated definitions and interpretations. Some scholars define privacy as the desire of people to have the freedom of choice under whatever circumstances and to whatever extent they expose their attitude and behavior to others [25]. Others state that privacy "represents the control of transactions between person(s) and other(s), the ultimate aim of which is to enhance autonomy and/or minimize vulnerability" [26, p.10]. Moreover, researchers have suggested that privacy is one's ability to control information about oneself [27], [28]. Different definitions and interpretations of privacy have led to the observation that there is a general lack of consensus on what privacy means. Because of the difficulties in defining privacy and also because the salient relationships depend more on cognitions and perceptions than on rational assessments, most empiric privacy research in the social sciences relies on the measurement of a privacy-related proxy of some sort [12].

In the information systems field for example, there has been a movement toward the measurement of IPC as the central construct between different antecedents and/or consequences [12]. Many studies have developed models with the aim of exploring the concept of IPC from data subjects in different contexts, including ecommerce, e-government, e-health, and social media. However, in the context of LA research in higher education, the IPC concept from the student perspective has been hitherto unexplored, and this study aims to fill this gap by proposing a relevant theoretical model.

## 2.2 Related studies

Unlike other contexts, to our knowledge, there is a lack of literature, related evidence as well as knowledge that pays attention to and explains students' IPC related to LA, in the context of higher education. The vast majority of studies that addressed privacy issues in this context have been largely focused on the institutional perspective rather than on the student perspective [7]. Few of the studies that attempted to cover this gap include the study of Ifenthaler and Schumacher [3]; they conducted a quasi-experimental study for surveying students' preferences for LA systems, their attitudes toward privacy as they relate to specific types of data, and how their attitudes influence the acceptance of LA systems. The results revealed that students consider some types of data (e.g., course enrollment, course results) to be much more important to keep private than others (e.g., medical data, income, marital status).

In another study, [28] carried out a survey aimed at investigating students' satisfaction with the practice of LA. Their findings showed students in particular were agreeable to the use of data to monitor their learning activity, and there was an overall acceptance of data usage to improve their grades. Jones et al. [29] conducted interviews to see how US students perceive privacy in relation to LA. The students highlighted that the process of data collection, sharing and usage should be clear to them. The study's findings stress that there is a general lack of awareness among students about the importance and functionality of LA.

Based on the views of students at the Open University (UK), Slade et al. [8] explored differences between students' attitudes to privacy and their online behaviors. The results indicated that there was a lower level of awareness about the collection, analysis and use of personal data. Also, the findings revealed that there is no obvious relationship between students' online frequency, privacy awareness and what they actually do to protect themselves. The findings also showed that students trust their university to use their data appropriately and ethically.

The study of Whitelock-Wainwright et al. [21] developed and validated the instrument SELAQ to measure 'ethical and privacy expectations' and LA 'service feature expectations' from the student perspective. The instrument was primarily validated from a UK university. One of the more recent studies [22] validated SELAQ in other three contexts – Estonian, Spanish, and Dutch universities – to assess how students' expectations may vary from one context to another. The findings show that the model provided acceptable fits in both the Spanish and Dutch context, but was not supported in the Estonian student context.

In another recent work, Botnevik [30] utilized several privacy principles (i.e., accountability, accuracy, anonymity, awareness, consent, data ownership, data security, data sharing, data preservation, limited access, opt-out, personal access, personal control, purpose, relevance, trust) as indicators to measure students' privacy perceptions in LA. The results suggest that the majority of students accept the use of LA but with data security and consent as the most important privacy principles for students. Also, Jones et al. [31] carried out an investigation to understand students' expectations of privacy issues related to LA, and their study showed that while students have high expectations of how the university handles their data, but they also held a self-protective attitude towards personal data.

All in all, a brief analysis of the aforementioned studies highlights that there is already some research literature in regards to the understanding of privacy issues and students' expectations and perceptions of LA, but there is still limited understanding of the concept of IPC. The majority of studies investigated students' awareness of privacy related issues, preferences and expectations, and overall, the focused related insights do not reveal students' concerns in relation to IPC. Thong and Hong [10] posit that perception of one's concern for others' behavior is different from one's expectation of others' behavior. From that, they also stress that IPC reflects an individual's perception of his or her concern for how personal information is handled by a certain institution, which is different from his or her expectation of how institutions should handle his or her personal information [10]. For example, an individual may expect an institution to provide adequate protection of his or her personal information, but it does not necessarily mean that this individual is genuinely concerned about providing his or her personal information to a specific agent. In the LA setting, "privacy expectations refer to how the university collects and analyses student data, specifically encompassing student expectations towards the provision of consent and the security of the data itself" [20]. This may imply that to investigate students' privacy expectations can help to reveal students' beliefs toward university privacy practices, but not necessarily individual beliefs or concerns that are based for example, on personal experience and privacy awareness gained from outside university boundaries.

## 3      Research Model

The present study aims at proposing a model that explores the concept of IPC from a student perspective in the LA context. To take one step further toward a more cohesive knowledge base that can guide a more responsible implementation of LA in higher education practice, we propose a model that is inspired by the APCO (antecedents–IPC–consequences) framework proposed by Smith et al. [31]. We have also adapted two antecedents; *perceived vulnerability*, and *perceived ability to control* from Dinev and Hart [16]. The motivation to start with these two constructs is grounded in the fact that these constructs account for concerns an individual can develop when determining whether to disclose personal information or not. These constructs "are related to a 'privacy calculus' that is, an assessment individuals make that their personal information will subsequently be used fairly and they will not suffer negative consequences"[16]. Further, the relationships between them and IPC are based on different interpretations and definitions of privacy per se [16]. Other previous studies (e.g., [10], [17]–[19]) have been also considered perceived vulnerability and/ or perceived ability to control as antecedents of IPC constructs.

Dinev and Hart [16] did not consider the consequences of IPC, but they suggested that future research may include other factors (e.g., trust) that may play an important role in mediating the control, vulnerability, and privacy concerns relationship. Other studies, including [10], [17], [19], [33] hypothesized that it is important to not only explore individual concerns or control-risks but to also explore how these concerns can be translated into trusting beliefs and information disclosure behaviors, that have also been found important in the learning analytics setting (Slade et al., 2019). Consequently, in our proposed model we include two more constructs (i.e., consequences of IPC), namely *trusting beliefs* and *self-disclosure behaviors*.

In summary, the research model (Figure 1) underlying the present research takes into account five constructs: the mediating variable; 1) information privacy concerns (IPC), two independent variables: 2) perceived vulnerability and 3) perceived ability to control, and two dependent variables: 4) trusting beliefs and 5) self-disclosure behavioral. For this study, we understand *perceived vulnerability* as "the perceived potential risk when personal information is revealed and has been considered in the literature as a factor that determines the perceived state of privacy and individual experiences" [16, p.415], *and perceived ability to control* as privacy control as the individual's beliefs in his or her ability to manage the release and dissemination of personal information [12], [16]. We consider *trusting beliefs* as the degree to which organizations (e.g., higher educational institutions) are dependable in protecting users' (e.g., students') personal information [18] and self-disclosure behaviors generally involve revealing information about oneself to others [34]. The relationship between these constructs resulted in 8 hypotheses.

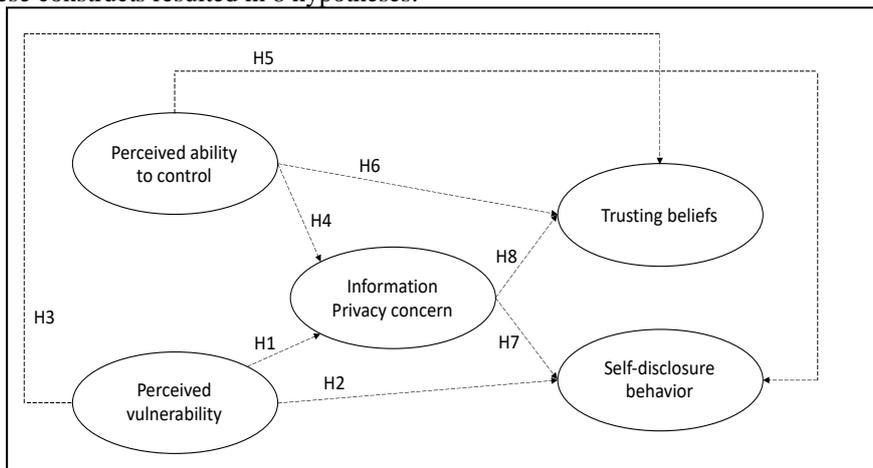

**Figure 1:** Research model

1. Increased perceived vulnerability increases IPC.
2. Increased perceived privacy vulnerability positively affects non-self-disclosure behavior.
3. Increased perceived vulnerability negatively affects trusting beliefs.
4. Increased perceived ability control reduces information privacy concerns.
5. Perceived ability to control reduces non-self-disclosure behavior.
6. Perceived ability to control increases trusting belief.
7. Information privacy concerns increase non-self-disclosure behavior.

8. Information privacy concerns reduce trusting beliefs.

Below, we present the theoretical foundation of the suggested hypotheses.

## 4 Research hypotheses

### 4.1 The effect of perceived vulnerability to information privacy concerns, trusting beliefs and self-disclosure behavior

The notion of vulnerability emerges from the complex definition of privacy. Perceived vulnerability describes the perceived potential risk when personal information is revealed and has been considered in the literature as a factor that determines the perceived state of privacy and individual experiences [16]. Various studies [12], [32]–[34] have shown that perceived vulnerability strongly influences individuals' IPC. Within the context of LA, students are exposed to a risk of misuse and abuse of personal information that might raise the perception of vulnerability e.g., [35]. They can be exposed to problems of surreptitious collection of, and unauthorized access to, their personal information that can be caused by many factors including insider curiosity or external threats [20]. All these factors contribute to increased perceived vulnerability on the students' side, and thus, their concerns related to information privacy[16]. Hence, we hypothesize that:

*H1: Increased perceived vulnerability increases IPC*

The perceived vulnerability is associated with risk that may affect an individual emotionally, materially, and physically [36]. In the LA setting, one risk that is associated with students' perceptions of privacy vulnerability is their awareness of the LAs' practices of collecting, analyzing, and reporting personal information. Consciously and unconsciously students may experience privacy vulnerability regarding the use of their personal data. This can deter students from sharing their personal information, perhaps even leading them to the provision of incomplete, false or inaccurate information [37]. Hence, we assume that students with high vulnerability perceptions should exhibit self-limitation toward disclosing personal information, such as refusing to give information to an institution because it is considered too personal. In line with this reasoning, we hypothesize that:

*H2: Increased perceived privacy vulnerability positively affects non-self-disclosure behavior.*

Malhotra et al. [18] define trusting beliefs as the degree to which organizations are dependable in protecting users' personal information. Others explain trusting beliefs as individual's beliefs that an organization (e.g., higher educational institution) will act according to their expectations, without exploiting their vulnerabilities [38]. To gain users'(students') confidence, organizations need to gain trust of the users so that if users were to allow organizations to access their personal information, this information will be safe and will not be exploited [39]. It has been also indicated that perceived privacy vulnerability influences the beliefs of trustworthiness. For example, Dinev and Hart [32] highlighted a direct negative effect of perceived vulnerability on individual trust on online services or institutions. Correspondingly, Liu et al. [40] suggest that the higher the perceived risks or vulnerability the lower level of trust. Consistent with them, we hypothesize that:

*H3: Increased perceived vulnerability negatively affects trusting beliefs.*

### 4.2 The effect of perceived ability to control to information privacy concerns, trusting beliefs and self-disclosure behavior

The ability to control individuals' information privacy is also embedded in the definition of privacy. Dinev & Hart [16] found that control is one of the major factors influencing privacy concerns. Other studies showed that low perceptions of control lead individuals to have high levels of IPC [41], and vice versa for high perceptions of privacy control [17]. Therefore, we hypothesize that:

*H4: Increased perceived ability to control reduces information privacy concerns.*

Previous studies argued that when control is allowed or when the future use of information is not known, individuals trust organizations [42], and consider that control is possible only through limiting self-disclosure [18], [43]. Hence, we hypothesize that:

*H5: Perceived ability to control reduces non-self-disclosure behavior,* and

*H6: Perceived ability to control increases trusting belief.*

### 4.3 The effect of information privacy concerns to trusting beliefs and self-disclosure behavior.

Researchers posit that people are more willing to disclose their personal information in online interactions if they perceive less information privacy concerns e.g., [33], [44], [45]. Moreover, Steward and Segars [46] also stress that individuals of high level of IPC are prone to non-self-disclosure behavior such as removing their names from mailing lists and refusing to provide personal information in the future. Consistent with this, we hypothesize that:

*H7: Information privacy concerns increase non-self-disclosure behavior.*

A consensus in the privacy-trust literature shows that individuals with high levels of IPC are likely to be low in trusting beliefs [18]. In addition, Thong and Hong [10] in their study, confirmed a significant negative effect of IPC to trusting beliefs. From that we also hypothesized that:

*H8: Information privacy concerns reduces trusting beliefs.*

## 5 Conclusion

In this study, we have proposed a theoretical model that aims to unveil students' *information privacy concerns* in regard to learning analytics in higher education. The concept of students' information privacy concerns is one of the key pillars in a more comprehensive understanding of students' privacy. A further theoretical justification as well as an empirical validation of the proposed model across different higher educational settings and cultures will assist in developing and practicing more effective practices at different levels aimed at protecting students' privacy in online higher education settings.